\documentclass[fleqn,usenatbib]{mnras}

\usepackage{amssymb}

\usepackage{newtxtext,newtxmath}
\usepackage{accents}
\newcommand{\ubar}[1]{\underaccent{\bar}{#1}}
\usepackage[T1]{fontenc}

\DeclareRobustCommand{\VAN}[3]{#2}
\let\VANthebibliography\thebibliography
\def\thebibliography{\DeclareRobustCommand{\VAN}[3]{##3}\VANthebibliography}

\usepackage{graphicx}	% Including figure files
\usepackage{amsmath}	% Advanced maths commands
\usepackage{amssymb}	% Extra maths symbols

\title[AGN in UNIONS post-mergers]{AGN in post-mergers from the Ultraviolet Near Infrared Optical Northern Survey}
\author[R. W. Bickley et al.]{Robert W. Bickley,$^{1}$\thanks{E-mail: rbickley@uvic.ca}
Sara L. Ellison,$^{1}$
David R. Patton,$^{2}$
Scott Wilkinson$^{1}$
\\
$^{1}$Department of Physics and Astronomy, University of Victoria, Victoria, British Columbia V8P 1A1, Canada\\
$^{2}$Department of Physics and Astronomy, Trent University, 1600 West Bank Drive, Peterborough, ON K9L 0G2, Canada\\
}

\date{Accepted XXX. Received YYY; in original form ZZZ}

\pubyear{2022}

\begin{document}
\label{firstpage}
\pagerange{\pageref{firstpage}--\pageref{lastpage}}
\maketitle

\begin{abstract}
The kinematic disturbances associated with major galaxy mergers are known to produce gas inflows, which in turn may trigger accretion onto the supermassive black holes (SMBH) of the participant galaxies. While this effect has been studied in galaxy pairs, the frequency of active galactic nuclei (AGN) in fully coalesced post-merger systems is poorly constrained due to the limited size or impurity of extant post-merger samples. Previously, we combined convolutional neural network (CNN) predictions with visual classifications to identify a highly pure sample of 699 post-mergers in deep $r$-band imaging. In the work presented here, we quantify the frequency of AGN in this sample using three metrics: optical emission lines, mid-infrared (mid-IR) colour, and radio detection of low-excitation radio galaxies (LERGs). We also compare the frequency of AGN in post-mergers to that in a sample of spectroscopically identified galaxy pairs. We find that AGN identified by narrow-line optical emission and mid-IR colour have an increased incidence rate in post-mergers, with excesses of \textasciitilde4 over mass- and redshift-matched controls. The optical and mid-IR AGN excesses in post-mergers exceed the values found for galaxy pairs, indicating that AGN activity in mergers peaks after coalescence. Conversely, we recover no significant excess of LERGs in post-mergers or pairs. Finally, we find that the [OIII] luminosity (a proxy for SMBH accretion rate) in post-mergers that host an optical AGN is \textasciitilde0.3 dex higher on average than in non-interacting galaxies with an optical AGN, suggesting that mergers generate higher accretion rates than secular triggering mechanisms.\end{abstract}

\begin{keywords}
galaxies: evolution -- galaxies: interactions -- galaxies: peculiar -- methods: statistical -- techniques: image processing
\end{keywords}

%%%%%%%%%%%%%%%%%%%%%%%%%%%%%%%%%%%%%%%%%%%%%%%%%%

%%%%%%%%%%%%%%%%% BODY OF PAPER %%%%%%%%%%%%%%%%%%

\section{Introduction}

Galaxy mergers are unique within the framework of hierarchical assembly in that they simultaneously transform the kinematics, morphologies, and intrinsic properties of the participant galaxies (\citealp{1978MNRAS.183..341W}; \citealp{1993MNRAS.262..627L}; \citealp{2008MNRAS.383...93B}; \citealp{2008ApJ...675.1095J}). Simulations (e.g. \citealp{1972ApJ...178..623T}; \citealp{Conselice_2006}; \citealp{2008MNRAS.391.1137L}) reproduce the observed signatures --- stellar shells, streams, and tails --- of interacting and post-merger galaxies (\citealp{2010MNRAS.401.1043D}; \citealp{2015ApJS..221...11K}; \citealp{2017MNRAS.464.4420S}). Numerous simulations are also in qualitative agreement about the chemical evolution and new star formation experienced by galaxies after they experience a disruptive, gas-rich merger (e.g. \citealp{1972ApJ...178..623T}; \citealp{2005MNRAS.361..776S}; \citealp{2008AN....329..952D}; \citealp{2019MNRAS.485.1320M}; \citealp{2020MNRAS.494.4969P}; \citealp{2020MNRAS.493.3716H}). Cold gas that might have been in a stable orbit in one or both progenitors can be disrupted by mergers, funneled towards the centres of the participant galaxies, and may be responsible for central starbursts suggested in observations (\citealp{2008AJ....135.1877E, 2013MNRAS.435.3627E}; \citealp{2004MNRAS.355..874N}; \citealp{2012MNRAS.426..549S}; \citealp{2014MNRAS.437.2137S}; \citealp{2015HiA....16..326K}; \citealp{2019MNRAS.482L..55T}). The cold gas that is responsible for central starbursts may also accrete onto a galaxy's super-massive black hole (SMBH) and produce the observational signatures of an active galactic nucleus (AGN; e.g. \citealp{2014MNRAS.441.1297S}; \citealp{Lackner_2014}; \citealp{10.1093/mnras/stw2620}; \citealp{2018PASJ...70S..37G}; \citealp{2011MNRAS.418.2043E, 2019MNRAS.487.2491E}). Together, increased star formation and energetic feedback from the AGN may enrich the circum-galactic medium through ejective means (\citealp{2015MNRAS.449.3263J}; \citealp{2018MNRAS.475.1160H}). The stellar and gas-phase kinematics of merger progenitors are also disrupted in the process (\citealp{1967MNRAS.136..101L}; \citealp{1977egsp.conf..401T}; \citealp{1983MNRAS.205.1009N}; \citealp{1992ApJ...400..460H}; \citealp{2003ApJ...597..893N}; \citealp{2006ApJ...645..986R}; \citealp{2009MNRAS.397.1202J}; \citealp{2014MNRAS.440L..66B}; \citealp{2018MNRAS.478.3994C}; \citealp{2018MNRAS.475.1160H}), and star formation may truncate rapidly due to either gas ejection or heating after the epoch of intense star formation and AGN feedback is complete (\citealp{1988ApJ...325...74S}; \citealp{2006ApJS..163....1H}; \citealp{2014ApJ...792...84Y}; \citealp{2021MNRAS.504.1888Q}). Indeed, \citet{2022MNRAS.517L..92E} recently identified an observational link between the post-merger phase and the signatures of rapid quenching using the same post-merger sample studied in \citet{2021MNRAS.504..372B} and this paper.

While a merger-AGN connection is therefore well established in pre-coalescence galaxy pairs, the precise role of post-coalescence mergers in switching on AGN and feeding them is not well constrained. Until recently, merger samples in the literature have been either too small to perform precise statistics, or dubious in purity. Without a large and highly pure merger sample, the quantitative role of mergers in SMBH evolution cannot be studied effectively.

Any effort to study the observed signatures of merger-induced phenomena across the entire merger sequence requires a post-merger sample that is both pure (containing as high a fraction as possible of genuine mergers) and representative. Merger identification is relatively straightforward in spectroscopic galaxy pairs, which can be identified by their visual appearances (\citealp{2007ApJS..172..329K}; \citealp{2005ApJ...625..621B}; \citealp{1998ApJ...499..112B}; \citealp{2010MNRAS.401.1043D}) or statistically, by grouping galaxies together in angular position and line-of-sight radial velocity in order to mitigate potential contamination by false pairs\footnote{Galaxies with small angular separations on the sky, but which are not destined or likely to merge on account of large separations in radial distance and/or velocity.}. Thanks to the large number of spectroscopically-identified galaxy pairs in redshift surveys, the statistical influence of the pair phase has already been explored in great detail (e.g., \citealp{2000ApJ...536..153P}; \citealp{2000ApJ...530..660B}; \citealp{2004ApJ...617L...9L}; \citealp{2005AJ....130.1516D}; \citealp{2008ApJ...681..232L}).

Small samples (e.g. \citealp{2013MNRAS.435.3627E}), and detailed spatially resolved case studies of individual post-mergers (e.g. \citealp{2019MNRAS.482L..55T}; \citealp{2015A&A...582A..21B}; \citealp{2019ApJ...881..119P}) offer a provisional understanding of post-coalescence galaxies. While these results have hinted at changes in star formation (e.g. \citealp{2013MNRAS.435.3627E}, \citealp{2020MNRAS.492.6027E}), chemical evolution (e.g. \citealp{2020MNRAS.494.3469B}), and intense SMBH activity (e.g. \citealp{10.1111/j.1365-2966.2011.20179.x}) in the post-merger epoch, fully coalesced galaxies are much more difficult to identify since they are no longer spectroscopically distinct from their companion(s). Consequently, the specific quantitative contribution of coalescence in producing these phenomena (especially SMBH triggering and accretion) is still being evaluated in simulations (e.g. \citealp{2022arXiv220314985S}) and observations (e.g. \citealp{2019MNRAS.487.2491E}; \citealp{2020A&A...637A..94G}).

Because the characteristic features of the post-merger phase are relatively faint, morphological merger identification methods require imaging of adequate depth and resolution (as demonstrated by \citealp{2019MNRAS.486..390B}, \citealp{2019MNRAS.489.1859H}, \citealp{2021arXiv211100961C}). The Canada France Imaging Survey (CFIS), part of the Ultraviolet Near Infrared Optical Northern Survey (UNIONS) collaboration, offers a useful combination of imaging quality and volume, with \textasciitilde0.7 arcsecond seeing, and $r$-band imaging that will eventually cover 5,000 square degrees of the sky. The survey's 5-$\sigma$ point-source depth (24.85 mag in the $r$-band for the MegaCam wide-field optical imager) is sufficient to capture the low-surface brightness features necessary for merger identification in bright, low-redshift galaxies (e.g. \citealp{2022arXiv220303973S}). Estimates of the low-$z$ merger rate suggest that the UNIONS footprint will include thousands of post-mergers (\citealp{1993MNRAS.262..627L}; \citealp{2011ApJ...742..103L}; \citealp{2012ApJ...747...34B}; \citealp{2014MNRAS.445.1157C}; \citealp{2015MNRAS.449...49R}; \citealp{2018MNRAS.480.2266M}).

Convolutional neural networks (CNNs) have already been successfully applied to a number of tasks in astronomy (e.g. \citealp{2015ApJS..221....8H}; \citealp{2018MNRAS.476.3661D}; \citealp{2019ApJS..243...17J}; \citealp{2019MNRAS.484...93D}; \citealp{2019ApJ...876...82N}; \citealp{2019MNRAS.489.1859H}; \citealp{2020ApJS..248...20H}), and are a natural candidate for merger identification in imaging (e.g. \citealp{2018MNRAS.479..415A}; \citealp{2019MNRAS.483.2968W}, \citealp{2019A&A...626A..49P}, \citealp{2020ApJ...895..115F}, \citealp{2020A&A...644A..87W}) and in stellar velocity fields (e.g. \citealp{2016ApJ...816...99H}, \citealp{2022MNRAS.515.3406M}, \citealp{2022MNRAS.511..100B}). The specific task of identifying a pure and complete post-merger sample in UNIONS imaging with a simulation-trained CNN is discussed in principle in \citet{2021MNRAS.504..372B}, and carried out in \citet{2022MNRAS.514.3294B}. After training and evaluation on realism-added mock CFIS observations of galaxies from the 100-1 run of the IllustrisTNG simulations (\citealp{2018MNRAS.480.5113M}; \citealp{2018MNRAS.477.1206N}; \citealp{2018MNRAS.475..624N}; \citealp{2018MNRAS.475..648P}; \citealp{2018MNRAS.475..676S}; \citealp{2019ComAC...6....2N}) the CNN was used to classify all CFIS galaxies with available SDSS Data Release 7 (DR7) spectra.

\citet{2021MNRAS.504..372B} noted that a CNN (or any automated merger classification method) with an accuracy of < 100\% (the CNN deployed in this work has an accuracy of \textasciitilde88\% on test set galaxies) would invariably fail to produce a pure post-merger sample on account of Bayesian statistics (\citealp{1763RSPT...53..370B}) and the minuscule prior probability that any given galaxy in the low-redshift universe will be a post-merger. To address this issue, \citet{2021MNRAS.504..372B} suggest that a hybrid method, in which a subset of galaxies predicted by the CNN to be post-mergers are subsequently inspected visually, would offer a reasonable combination of efficiency and reproducibility without sacrificing the purity of the final post-merger sample. The strengths of CNNs and human classifiers are complementary: a simulation-trained CNN can classify galaxies quickly, rule out a large number of galaxies unlikely to be mergers, and capture a breadth of observed merger characteristics spanning the range of remnant stellar masses, merger mass ratios, redshifts, orbital parameters, and observational conditions included in the training set. Conversely, human classifiers are capable of meticulous classification with the goal of sample purity in mind, and are capable of explaining and defending their decisions.

As long as the CNN's training data is observationally realistic (e.g. \citealp{2019MNRAS.490.5390B}; \citealp{2019MNRAS.489.1859H}; \citealp{2021arXiv211100961C}) and the simulated galaxies exhibit the same morphological characteristics as galaxies in the low-redshift universe (e.g. \citealp{2019MNRAS.483.4140R}; \citealp{2021MNRAS.501.4359Z}), visual classifiers ought to inherit a sample with a high post-merger fraction from the CNN. Critically, this particular combination of classifiers also improves over previous post-merger identification efforts in the diversity of post-mergers included in the final sample. If less disturbed post-merger morphologies are included in the CNN's training set and given appropriate consideration in the visual classification phase of the hybrid method, they can be preserved and studied alongside more visually obvious major mergers as long as their disturbed morphologies are sufficiently bright to be captured at the depth of CFIS. Methods using shallower (e.g., SDSS) imaging, visual classifications (e.g., \citealp{2013MNRAS.435.3627E}) or CNNs trained on visual classifications (e.g., \citealp{2019A&A...631A..51P}; \citealp{2020A&A...637A..94G}) do not reap the same benefit --- even when galaxies are inspected with great care, the most dramatic merger morphologies always inspire more confidence in visual classifications.

In this work, we briefly review the simulation-trained CNN and our method for its deployment in a hybrid classification scheme, as well as three observational methods (optical spectroscopy, mid-IR photometry, and radio classification) of AGN identification (Section~\ref{Methods}). We next use the visually confirmed post-merger galaxies from \citet{2022MNRAS.514.3294B} to investigate the link between the merger sequence and the triggering of each AGN type (Section~\ref{AGN in post-mergers}), and investigate the reasons for differences between the \citet{2013MNRAS.435.3627E} post-merger results and our own. Finally, we select post-merger and SDSS pair samples of optical AGN, and compare their [OIII] luminosities with those of control AGN in order to approximate the accretion rate enhancements produced by ongoing and completed mergers. We assume cosmological parameters ($\Omega_{\mathrm{m0}}$ = 0.3, $\Omega_{\Lambda0}$ = 0.7, $h$= 0.7) when calculating luminosity distances, and for any other cosmology-dependent quantities appearing in this work.

\section{Methods}
\label{Methods}

\subsection{UNIONS data}
\label{UNIONS data}

Our census of AGN in post-mergers would not be possible without high quality imaging resolution and depth over a large sky area. The UNIONS collaboration is a new consortium of wide field imaging surveys of the northern hemisphere and represents an excellent opportunity for merger searches. UNIONS consists of 1) CFIS conducted at the 3.6-meter CFHT on Maunakea, 2) members of the Pan-STARRS team, and 3) the Wide Imaging with Subaru HyperSuprimeCam of the Euclid Sky (WISHES) team. \citet{2021MNRAS.504..372B} used only the $r$-band imaging in creating synthetic images, and the CNN predictions used in this work are made exclusively on CFIS $r$-band imaging.

The observing pattern employed by CFIS uses three single-exposure visits with field-of-view (FOV) offsets in between for optimal astrometric and photometric calibration with respect to observing conditions. This also ensures that the entire survey footprint, including areas in the "chip gaps" between MegaCam's multiple CCDs for a given exposure, will be visited for at least two exposures. After raw images are collected by CFHT, they are detrended (i.e., the bias is removed and the images are flat-fielded using night sky flats) with the software package MegaPipe (\citealp{2008PASP..120..212G,2019ASPC..523..649G}). The images are next astrometrically calibrated using Gaia data release 2 (\citealp{2016A&A...595A...1G,2018A&A...616A...1G}) as a reference frame. Pan-STARRS 3$\pi$ $r$-band photometry (\citealp{2016AAS...22732407C}) is used to generate a run-by-run differential calibration across the MegaCam mosaic, and an image-by-image absolute calibration. Finally, the individual images are stacked onto an evenly spaced grid of 0.5-degree-square tiles using Pan-STARRS PS1 stars as in-field standards for photometric calibration. The resulting $r$-band images have a typical 5-$\sigma$ point-source depth of 24.85 mag, \textasciitilde0.6-arcsecond seeing, and a pixel size of 0.187 arcseconds.

In order to prepare CFIS galaxy images for classification, we cropped them to a physical scale of 100 kpc on a side based on their redshifts, resized to 138 $\times$ 138 pixels, and normalized on a linear scale with brightnesses of 0 and 1 assigned to the faintest and brightest pixels, respectively. Since this method requires an accurate spectroscopic redshift in order to crop and scale each galaxy image appropriately, we restrict our post-merger search to galaxies with $z$ < 0.5 observed by both the CFIS data release 2 (DR2) and SDSS DR7, and match the two catalogs with a 2-arcsecond tolerance, yielding 168,597 galaxies. Appearances of the same galaxies in the archival Wide-field Infrared Survey Explorer (WISE) space telescope\footnote{wise2.ipac.caltech.edu/docs/release/allsky/} photometric catalog as well as the NRAO VLA Sky Survey (NVSS) and Faint Images of the Radio Sky at Twenty-centimeters (FIRST) radio catalogs will facilitate subsequent characterization of their SMBHs; see Sections~\ref{Infrared AGN} and~\ref{LERGs}. After processing, the galaxy images have the same physical and pixel dimensions as the synthetic images used for training in \citet{2021MNRAS.504..372B}.

The $r$-band imaging data is supplemented by derived measurements (e.g., redshifts) from the MPA-JHU catalog of derived properties\footnote{wwwmpa.mpa-garching.mpg.de/SDSS/DR7/} from SDSS DR7, stellar mass estimates from \citet{2003MNRAS.341...33K}, and corrected optical emission line fluxes from \citet{2012MNRAS.426..549S}. In Section~\ref{WISE AGN excess} we use non-parametric morphological statistics from \citet{2016MNRAS.456.3032P} to characterize degrees of merger-induced disturbance.

\subsection{The visually confirmed post-merger sample}
\label{The visually confirmed post-merger sample}
% CNN deployment and visual classifications

In \citet{2021MNRAS.504..372B}, the CNN was trained on realism-added images of IllustrisTNG post-merger remnants with mass ratios of at least 1:10, stellar masses between $\mathrm{10^{10-12}}$ \(\textup{M}_\odot\), and which had experienced coalescence between the previous and present simulation snapshots (corresponding to a timescale of $\leq$160 Myr) every snapshot since simulation $z$=1, and control galaxies that had not experienced a merger in at least 2 Gyr of simulation time. \citet{2021MNRAS.504..372B} demonstrated that the network classified reserved simulated post-mergers and control images with \textasciitilde88\% accuracy, but cautioned that high accuracy would not directly lead to a pure predicted post-merger sample on account of the rarity of post-mergers in the real Universe (and simulated universes as well). As a result, \citet{2021MNRAS.504..372B} explored the utility of progressive cuts in CNN p(x), or the floating point prediction between 0$-$1 returned by the network as a prediction of merger status. In a sample with a realistic proportion of post-mergers, using p(x) > 0.5 to select mergers would inevitably lead to a sample of inadequate purity. \citet{2021MNRAS.504..372B} argued that a hybrid approach, in which a more stringent p(x) cut is combined with subsequent visual inspection, could be an efficient way forward.

Our CNN was first used to classify real CFIS galaxies in \citet{2022MNRAS.514.3294B}, and the same classifications are prerequisite to this work. The CNN assigned p(x) predictions > 0.5 to 6,778 galaxies. Because of the natural rarity of post-merger galaxies, visual inspection showed the galaxy sample with p(x) > 0.5 to be significantly contaminated with non-interacting galaxies, a result consistent with our expectations based on Bayesian statistics (see Section 3.6 in \citealp{2021MNRAS.504..372B}). We therefore chose to visually inspect the 2,000 galaxies with classifications greater than p(x) > 0.75. There is no special significance to p(x) > 0.75 other than the fact that it produces a reasonable number of predicted post-mergers for subsequent inspection by a classification team. In order to identify a pure post-merger sample, we used a conservative classification method that required a consensus post-merger classification with mutual agreement by the authors RWB, SLE, and DRP after careful inspection of the CFIS galaxy image, the wider field, and spectroscopic companions via SDSS. Galaxies for which a consensus could not be reached were discarded. Of the 2,000 galaxies inspected, 1,201 were rejected by the classification team on account of doubtful merger status, and a consensus was not reached for 100 additional galaxies. After visual inspection, a sample of 699 galaxies with unambiguous post-merger morphologies (described fully in \citealp{2022MNRAS.514.3294B}) were confirmed unanimously by all three authors.

The merger sample in \citet{2010MNRAS.401.1043D} was selected based on the merger vote fraction of citizen scientist volunteers on SDSS imaging, and consisted of 3003 visually selected mergers (although not all are post-coalescence). \citet{2013MNRAS.435.3627E} inspected 370 galaxies from the \citet{2010MNRAS.401.1043D} merger catalog which were strongly perturbed but lacked a spectroscopic companion (suggesting either a flyby or a completed merger) by eye and selected a smaller sample of 97 post-mergers. Our post-merger catalog, first presented in \citet{2022MNRAS.514.3294B}, offers an improvement in post-merger purity over \citet{2010MNRAS.401.1043D} and a substantial increase in size compared to \citet{2013MNRAS.435.3627E}. The post-mergers studied in this work were also selected using deeper imaging, and as a result include fainter and more subdued merger morphologies than could be detected using SDSS imaging.

\begin{figure*}
\includegraphics[width=\textwidth]{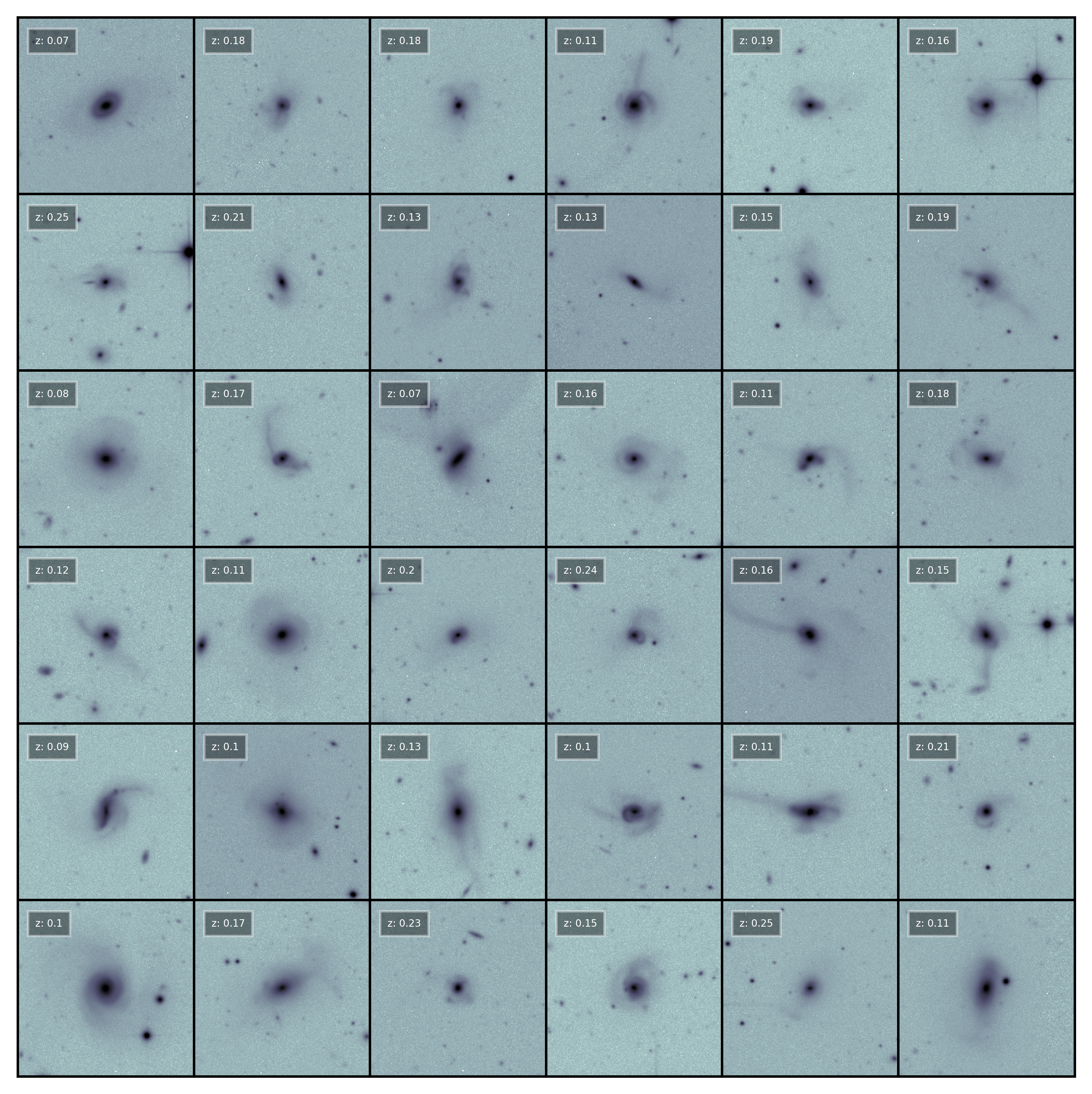}
\caption{A mosaic of 36 post-mergers from our hybrid (CNN-identified and visually confirmed) post-merger sample, cropped to a physical size of 100 kpc on a side. All galaxies in the sample have distinctive merger-induced morphologies that could not be plausibly produced by any discernible object in the CFIS imaging, or by any spectroscopic companions as identified using SDSS. Furthermore, each of the post-mergers has only a single post-coalescence bright nucleus. Galaxies are shown in log-scale with the contrast adjusted for consistency.}
\label{fig:pm-mosaic-36}
\end{figure*}

36 post-mergers from this visually confirmed sample are shown with logarithmic brightness scaling in Figure~\ref{fig:pm-mosaic-36}. As with all confirmed post-mergers in the sample, the galaxies exhibit unambiguous merger-induced morphologies: tidal tails, streams, or shells. While absolute knowledge of the status of each galaxy's nucleus is limited by the resolution and seeing of the imaging, each galaxy in the sample appears to have only a single bright nucleus. In addition, we ruled out any galaxies whose merger-like features could plausibly have been induced by any nearby object imaged by CFIS, or identified by projected separation in SDSS. By eliminating companions both near and far, we can confidently deduce that the merger features in each of our 699 galaxies were produced by a partner that has been wholly accreted at the time of imaging.

The study of star formation in the visually confirmed post-merger sample in \citet{2022MNRAS.514.3294B} lent credence to the suggestion in \citet{2021MNRAS.504..372B} that a hybrid approach to post-merger identification, starting with a preliminary filtering by the CNN and ending with rigorous human visual classifications, would be an efficient way to identify a pure sample of galaxies belonging to the exceedingly rare post-merger class in a large observational sample. Even though the visually confirmed sample is not free of  biases, we maintain that the \citet{2022MNRAS.514.3294B} visually confirmed post-merger sample is likely representative of the remnants of major mergers.

\begin{figure}
\includegraphics[width=\columnwidth]{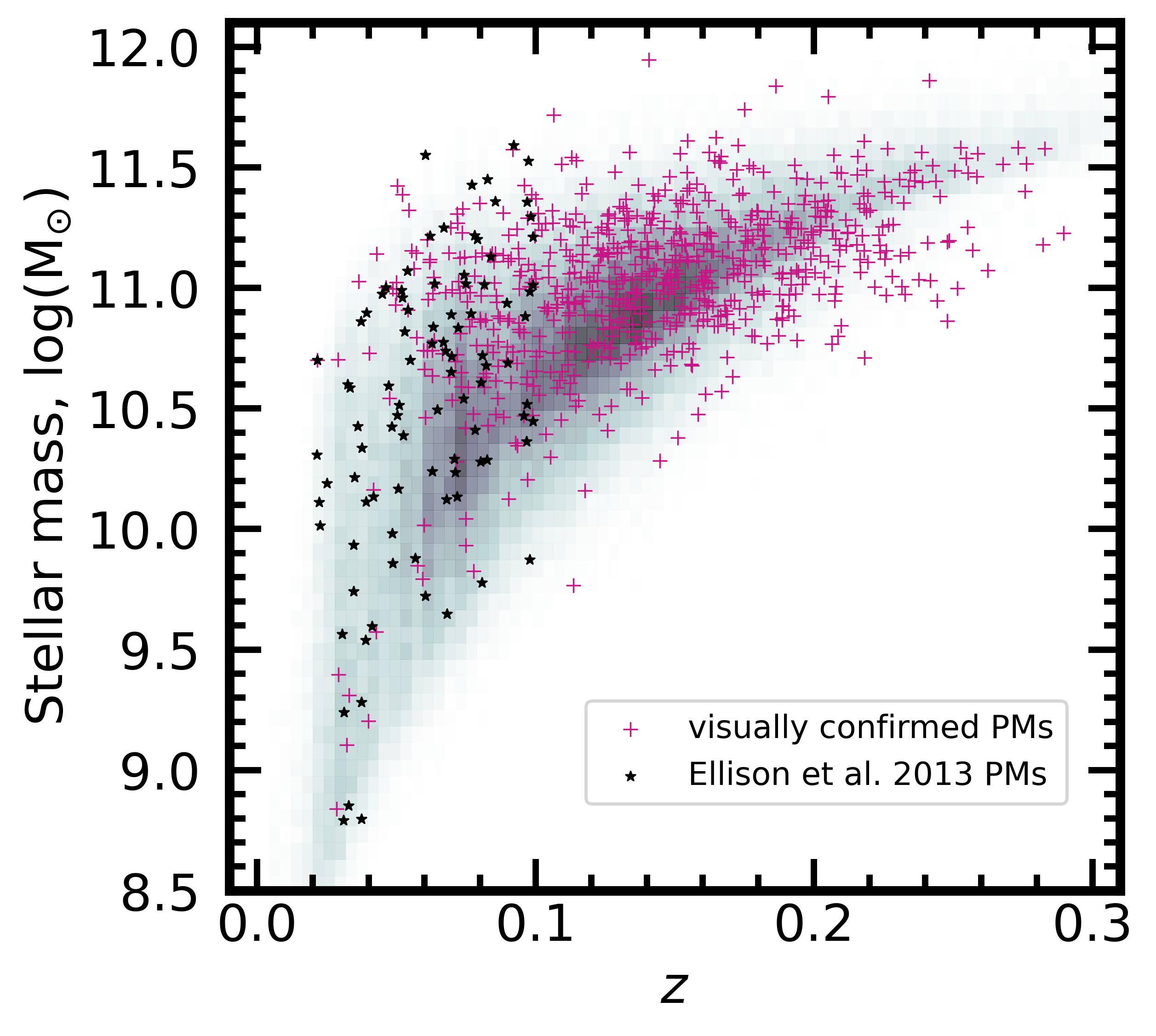}
\caption{The stellar masses and redshifts of the 699 visually confirmed post-merger galaxies (magenta crosses) studied in the present paper and 97 \citet{2013MNRAS.435.3627E} post-mergers (black stars) superimposed over CFIS DR2 parent sample (gradient histogram). CFIS $r$-band imaging was processed and used as the input for CNN classification as well as for visual inspection in \citep{2022MNRAS.514.3294B}.}
\label{fig:m-z-dist}
\end{figure}

We also compare the characteristics of the visually confirmed, larger post-merger sample to those of the \citet{2013MNRAS.435.3627E} post-mergers, which were visually selected from the larger \citet{2010MNRAS.401.1043D} post-merger catalog by the author SLE. Though lesser in number (97 in the sample before any experiment-specific cuts are applied), they provide useful context: they represent a highly pure sample as well, albeit one that is likely biased towards mergers of greater visual strength due to their selection from shallower SDSS imaging.

Figure~\ref{fig:m-z-dist} shows the stellar masses and redshifts of both post-merger samples compared to the parent sample of galaxies from CFIS DR2. Both merger samples span the dynamic range of CFIS DR2 galaxy stellar masses, but the new visually confirmed post-merger sample identifies more distant post-merger galaxies out to $z$\textasciitilde0.3, whereas the \citet{2013MNRAS.435.3627E} post-mergers lie below $z$ < 0.1. The success of the CNN at identifying galaxies over the full CFIS redshift range likely stems from its training set, which was composed of simulated galaxies inserted at redshifts drawn at random from the real CFIS distribution. The visually confirmed post-mergers also preferentially lie at high stellar masses compared to the parent sample. This bias is discussed at length in \citet{2022MNRAS.514.3294B}, and we posit two candidates for the cause. First, the CNN was trained on galaxies with masses $\mathrm{M_{\star}}$ > $\mathrm{10^{10}}$ \(\textup{M}_\odot\) in order to ensure that their morphologies were adequately resolved by the simulation, and it is natural for the CNN  to preferentially select galaxies similar to those on which it was trained. Second, the observability of merger-induced morphologies depends on the brightness of a galaxy relative to that of the sky, and more massive galaxies are likely to be brighter in appearance. This bias was affected during the CNN phase of the \citet{2022MNRAS.514.3294B} hybrid merger identification effort, and therefore propagates through to the visually confirmed post-merger sample as well.

\subsection{Galaxy pairs}
\label{Galaxy pairs}

Though this work only addresses the AGN characteristics of observed post-mergers, the characteristics of the post-merger epoch of galaxy evolution is best contextualized within the merger sequence. A sample of pre-coalescence galaxy pairs is therefore required. We use the galaxy pair sample compiled by \citet{2016MNRAS.461.2589P}, and select objects for comparison emulating the method employed in \citet{2013MNRAS.435.3627E}. Whenever they are shown for context in this work, SDSS pairs are required to have projected separations $r_{p}\leq80h_{70}^{-1}$ kpc, line of sight velocity differences $\mathrm{\Delta}$V $\leq$ 300 km s$^{-1}$ (in order to minimize accidental projections of a false pair), and stellar mass ratios of $0.1 \leq$ M$_{1}$/M$_{2} \leq 10$.

Note that we are applying symmetrical mass ratio criteria to our pair samples and the post-merger samples used to train our CNN (1:10 or more, see Section~\ref{The visually confirmed post-merger sample}). Although merger mass ratio may scale with the degree of morphological disturbance of merger remnants, we do not find that the inclusion / exclusion of minor galaxy pairs (e.g., changing the mass ratio criterion to $0.25 \leq$ M$_{1}$/M$_{2} \leq 4$) has a significant influence on the excesses we calculate. We also expect the median progenitor mass ratio of our post-merger sample to increase following the visual inspection effort detailed in \citet{2022MNRAS.514.3294B}, since merger candidates of unconvincing strength were removed. We therefore present the physical characteristics of merger remnants alongside galaxy pairs with the caveat that the pairs studied may not be precise pre-merger analogues to the post-mergers.

A larger $\mathrm{\Delta}$V requirement, e.g., $\leq$ 1000 km s$^{-1}$ could capture a larger sample of galaxies that could conceivably be experiencing interactions, but the risk of contamination by non-interacting galaxies also increases. In order to restrict our analysis to galaxies that are very likely to be pre-mergers, we therefore follow \citet{2016MNRAS.461.2589P} and use $\mathrm{\Delta}$V $\leq$ 300 km s$^{-1}$. Regardless of the choice of $\mathrm{\Delta}$V, we expect some contamination from interlopers to persist.

\subsection{Control pools}
\label{Control pools}

Wherever post-mergers are studied in this paper, we will compare their physical characteristics to those of control galaxies that are similar in mass and redshift, but which are not themselves post-mergers. To this end, we again use the \citet{2022MNRAS.514.3294B} CNN classifications to our advantage. The vast majority (131,168 of 168,597) of galaxies in the CFIS sample are assigned CNN p(x) < 0.1 by the network, nominally indicating non-post-merger status. Even with an accurate classifier, Bayesian statistics suggest that a small number of genuine post-mergers will fall below this threshold. Thanks to the initial rarity of post-mergers, however, we can reasonably assume that the extreme dilution of post-mergers below p(x) < 0.1 will effectively eliminate their influence on our results.

Wherever the \citet{2013MNRAS.435.3627E} sample is shown, their physical characteristics are compared to those of their own mass- and $z$-matched corresponding control sample with a Galaxy Zoo merger vote fraction of zero. As with the visually confirmed post-merger sample, we expect that the signal from misclassified post-mergers contaminating this control sample should be effectively zero.

The controls for galaxy pairs must have projected separations of $r_{p}>80h_{70}^{-1}$ kpc, and Galaxy Zoo (\citealp{2010MNRAS.401.1043D}) merger vote fractions of zero in order to ensure that they do not belong to an interacting pair, and have not merged recently. The particular control matching methodologies for each of our experiments are described in detail in Section~\ref{Results}.

\subsection{Optical (Seyfert II) AGN}
\label{Optical AGN}
%SDSS measurements, AGN criteria, AGN accretion rates & justifications for all

Since spectroscopic redshifts are required for appropriate treatment of our CFIS galaxy images, all post-mergers identified in CFIS $r$-band imaging by the CNN have available SDSS DR7 optical spectra as well. Conveniently, we can use emission line measurements from these same optical spectra to search for the luminous echoes of AGN in our galaxies' gaseous narrow-line regions (NLR). Galaxies exhibiting the NLR optical emission line characteristics of the Seyfert II class will hereafter be referred to as "optical AGN".

In order to quantify the incidence of optical AGN in post-mergers, as well as estimate the accretion rates of their black holes, we use the "maximum starburst line" on the Baldwin, Phillips, Terlevich (BPT; \citealp{1981PASP...93....5B}) diagram computed by \citet{2001ApJ...556..121K}. The line represents the theoretical boundary separating AGN from starbursts, determined using grids of models with varied metallicities and ionization parameters, and is more stringent than the criterion of \citet{2003MNRAS.346.1055K} which is often used to separate purely star-forming galaxies from those with potential contributions from AGN. We use Milky Way and host extinction-corrected optical emission line measurements (from the SDSS MPA-JHU catalog and \citealp{2012MNRAS.426..549S}) to place our target and control galaxies on the BPT diagram. In order to avoid contamination from shocks, which are expected in gas-rich galaxy mergers, we do not count objects on the AGN side of the \citet{2001ApJ...556..121K} diagram as AGN if they fall below either the [SII] or [OI] BPT diagram low-ionization nuclear emission region (LINER) criteria described in \citet{2006MNRAS.372..961K}.

When quantifying AGN excesses (the ratio of the AGN fractions in mergers and controls) in Section~\ref{AGN in post-mergers}, we require S/N $\geq 5$ on all four emission lines used in the BPT diagram. Consequently, the nebular emission lines of galaxies that we count in our excess calculations are dominated by the high-energy ionization associated with the AGN, and our excess calculations capture the frequency with which mergers induce strong, unambiguous optical evidence of AGN.

Later, when we approximate the SMBH accretion rate with the luminosity of the [OIII] emission line in Section~\ref{Optical AGN accretion rate enhancements}, we require an ensemble of 5 or more controls (which must also be optical AGN) for each merger in order to calculate a robust accretion rate enhancement. The [OIII] emission line has been used many times to approximate the accretion rates of optically identified AGN, (e.g. \citealp{2003MNRAS.346.1055K}; \citealp{2004MNRAS.351.1151B}; \citealp{2009ApJ...695L.130C}; \citealp{2012ApJ..745...94L}), but [OIII] flux can also be contaminated by star formation. The link between [OIII] luminosity and star formation rate is complex, and its dependence on several parameters is explored in \citet{2001ApJ...556..121K}. Nonetheless, the \citet{2001ApJ...556..121K} criterion we use is intended to be a maximal starburst division, such that galaxies above it can only be produced (in their models) by AGN contributions. We choose galaxies for this experiment whose emission is unambiguously AGN-dominated according to the same \citet{2001ApJ...556..121K} criterion used in our excess calculations, and compare the AGN found in mergers to those found in non-mergers. The novel size of the visually confirmed post-merger sample allows us to use the same S/N $\geq 5$ cut as we do when measuring optical AGN excesses, and additionally remove LINERs using the [SII] and [OI] \citet{2006MNRAS.372..961K} criteria in order to ensure a robust connection between observed [OIII] luminosity and SMBH accretion.

\subsection{Infrared AGN}
\label{Infrared AGN}
%WISE survey, AGN criterion & justification

Though they are useful for identifying a particular subset of AGN, optical emission lines do not represent a complete census of SMBH activity in low-redshift galaxies (\citealp{2018ARA&A..56..625H}). A number of studies have demonstrated the utility of mid-infrared (mid-IR) observations to identify dust-obscured AGN (e.g. \citealp{2004ApJS..154..166L}; \citealp{2005ApJ...631..163S}; \citealp{2007ApJ...660..167D}; \citealp{2007ApJ...671.1365H}; \citealp{Donley_2008}; \citealp{Eckart_2009}; \citealp{2012ApJ...753...30S}; \citealp{2013MNRAS.434..941M}). Many galaxies with ongoing SMBH accretion are obscured by dust such that their optical emission line strengths cannot be reliably measured, while at the same time reddening their mid-IR colours (e.g. \citealp{2013ApJ...772...26A}). In fact, mergers may preferentially produce dust-obscured AGN (\citealp{2014MNRAS.441.1297S}, \citealp{2017MNRAS.464.3882W}, \citealp{2018MNRAS.478.3056B}). Models predict that when one or both companions involved in the interaction is dusty, the kinematic disturbances induced by the final stages of a merger may distribute the dust in the centre, obscuring optical emission that might be observable in a dynamically settled galaxy (\citealp{2022arXiv220500567Y}).

In order to identify dust-obscured AGN, we turn to legacy all-sky mid-IR observations from the WISE space telescope, and follow \citet{2014MNRAS.441.1297S} and \citet{2013MNRAS.435.3627E} in deploying a WISE W1$-$W2 > 0.5 colour cut. Because we are interested in studying the proportion of galaxies with WISE photometry whose W1$-$W2 colours exceed 0.5, we also require that all galaxies (post-mergers as identified by any method, galaxy pairs, and controls) included in our WISE-related results have been detected by WISE. WISE sources considered in our analysis have S/N of at least 2 for both W1 and W2, but a more stringent S/N criterion does not change our results since > 99\% of our sources have S/N > 10 for both W1 and W2. Mid-IR observations can be affected by star formation, but even template spectra for galaxies with extreme star formation have been shown to fall below W1$-$W2 = 0.5 at low $z$ (\citealp{2013ApJ...772...26A}; \citealp{2014MNRAS.441.1297S}). Still, our visually-confirmed post-mergers are known to be enhanced in star formation (\citealp{2022MNRAS.514.3294B}), so it remains conceivable that a small contribution from star formation may affect the signal in Section~\ref{WISE AGN excess}. Caution should therefore be exercised in direct interpretation of the quantities reported therein.

\subsection{LERGs}
\label{LERGs}
%Best radio catalog, LERG criterion & justification

Optical and mid-IR AGN detections together provide a reasonably complete census of the observational phenomena associated with radiatively efficient, rapidly accreting SMBH with a luminous accretion disk. Indeed, these AGN are most closely linked with the archetypal understanding of the role of the merger sequence in galaxy evolution, in which gas inflows simultaneously produce upticks in star formation and SMBH accretion (e.g. \citealp{2006ApJS..163....1H}). While useful, this narrative largely excludes mergers between gas-poor systems. Although low-excitation AGN states are most often associated with relatively isotropic accretion of hot gas from galactic halos (e.g. \citealp{2005MNRAS.362...25B,2006MNRAS.368L..67B}; \citealp{2006MNRAS.372...21A}; \citealp{10.1111/j.1365-2966.2007.11572.x}; \citealp{10.1093/mnras/stt692}), there has been some evidence of a role for mergers in triggering radiatively inefficient accretion (e.g. \citealp{10.1093/mnras/sts675}; \citealp{2019MNRAS.489.2308G}).

In order to investigate the role of mergers in this context, we also quantify the proportion of low-excitation radio galaxies (LERGs) in our merger and pair samples relative to controls selected from the same SDSS parent sample as in Section~\ref{Optical AGN}. These low-excitation radio-selected AGN are taken from the compilation of \citet{10.1111/j.1365-2966.2012.20414.x}, who match SDSS to a pair of radio catalogs: NVSS and FIRST. After classifying their detections as either star forming or AGN via a combination of 4000\AA~break strengths, radio-to-emission-line luminosities, and BPT diagnostics, they use optical emission lines and a second decision tree to distinguish the latter into high-excitation radio galaxies (HERGs) and LERGs. Because the high-excitation AGN state is well described by our optical and mid-IR observations, we consider only the LERGs in this work.

\subsection{Overlap of AGN types}
\label{Overlap}
The visually confirmed post-merger sample, \citet{2013MNRAS.435.3627E} post-merger sample, and SDSS galaxy pair sample contain 699, 96, and 17,566 galaxies, respectively. The visually confirmed post-merger sample contains the following:

\begin{itemize}
\item 32 optical AGN meeting the \citet{2001ApJ...556..121K} and \citet{2006MNRAS.372..961K} criteria (excluding star-forming, composite, and LINERs) with S/N > 5 for the emission lines used in BPT placement
\item 66 mid-IR AGN with W1$-$W2 > 0.5
\item 14 LERGs
\item 17 galaxies that host AGN identifiable using both optical and mid-IR criteria
\item No overlap between either the optical or mid-IR AGN and LERGs
\end{itemize}

On average, the optical AGN have a WISE colour of \textasciitilde0.35, and the LERGs have a typical WISE colour of \textasciitilde0.12. Our choices of criteria for optical and mid-IR AGN do not by definition preclude the possibility of a galaxy being a LERG, and such dual-status objects do appear in the SDSS DR7 and WISE catalogues. They are rare, however, and as a result none appear in the visually confirmed post-merger sample.

\section{Results}
\label{Results}

\subsection{AGN in post-mergers}
\label{AGN in post-mergers}

After identifying our post-merger samples and establishing AGN criteria, we can quantify the strength of the link between merger status and SMBH incidence by calculating AGN excesses --- the ratio of the AGN fractions in matched samples of post-mergers (or pairs) and controls. When controls are properly matched and statistics are of adequate quality, any excess > 1 indicates that the differentiating characteristic between the two samples (i.e., post-merger or pair status) is responsible for (or correlated with a factor that is responsible for) the increased AGN frequency.

\begin{figure}
\includegraphics[width=\columnwidth]{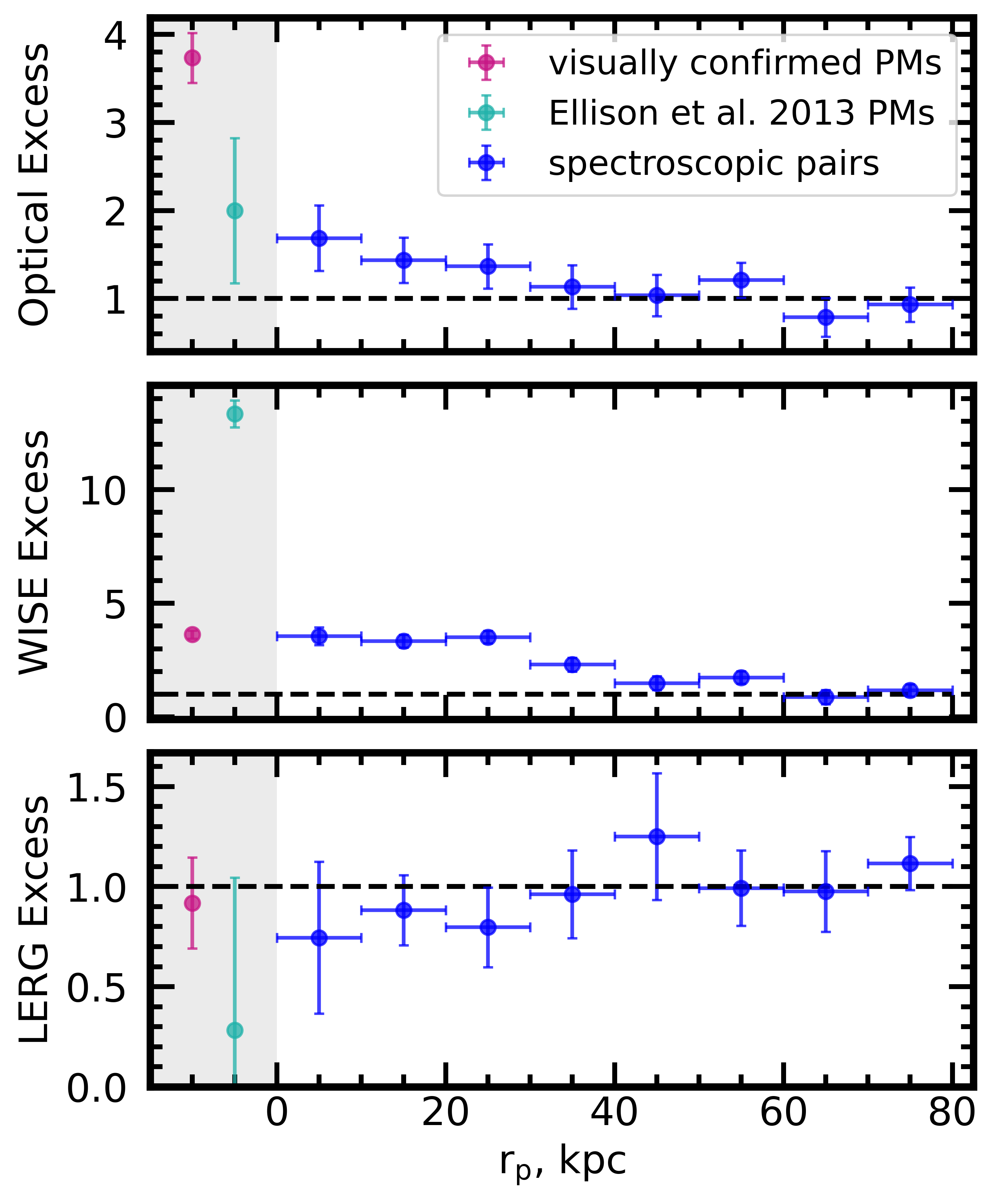}
\caption{Optical (top), mid-IR (centre), and LERG (bottom) AGN excesses in SDSS spectroscopic pairs (blue), the \citet{2013MNRAS.435.3627E} post-merger sample (teal), and the visually confirmed post-merger sample (magenta). Vertical errors are calculated by adding the inverses of the binomial errors on each fraction $\sqrt{f(1-f)/N}$, where f is the AGN fraction in the target or control sample, and N is the size of that sample. Horizontal errors are the bin widths. Our visually confirmed, more inclusive post-merger sample finds a stronger optical AGN excess, and a weaker mid-IR AGN excess. This is consistent with the hypothesis that more minor mergers, which are present in the visually confirmed sample and largely absent from the \citet{2013MNRAS.435.3627E} sample, are less likely to be dust-obscured after the merger. We find LERG excesses for post-mergers and galaxy pairs consistent with unity. Even with a relatively simple control-matching scheme, the merger sequence does not appear to be strongly connected to LERG status.}
\label{fig:exc-mz}
\end{figure}

In calculating each excess, we match controls to the post-mergers or pairs (hereafter referred to collectively as "target galaxies") on stellar mass and redshift. Controlling for environment using additional parameters (halo mass, environment density) does not qualitatively change our results, suggesting that the influence of the most relevant interacting companion (in the case of galaxy pairs) or coalesced companion (for the post-mergers) is largely responsible for the signal uncovered in this Section. In order to ensure we have the same number of controls per target galaxy, we identify the closest non-merger match (without replacement) in 2-D parameter space for each target galaxy, assigning equal weight to each parameter. After finding one control for each galaxy, a Kolmogorov–Smirnov (K-S; \citealp{smirnov1948}) statistic is calculated for the M$\star$ and $z$ distributions of the target galaxies and controls. If the K-S $p$ value is below 0.9, the control pool is finalized and we calculate the AGN fractions and excess for the relevant observational type (optical, mid-IR, and LERG). If the $p$ value remains above 0.9, we attempt to match additional equal-sized batches of controls, checking the K-S $p$ value every time before adding an additional batch. We cap the number of controls per post-merger at 10, as the counting statistics cease to improve significantly by that time. We use the same statistical control matching methodology for each target sample (\citealp{2022MNRAS.514.3294B} post-mergers, \citealp{2013MNRAS.435.3627E} post-mergers, and galaxy pairs), allowing for the control pool (see Section~\ref{Control pools}) and number of controls per target galaxy to vary (between 1$-$10 depending on the continued success of K-S tests) between each subset of galaxy pairs (binned by projected separation) and post-merger sample we study.

In the target samples and the matched control samples, the optical AGN fraction is the number of BPT AGN above the \citet{2001ApJ...556..121K} and \citet{2006MNRAS.372..961K} lines with S/N $\geq 5$ divided by the total number of target or control galaxies. The WISE AGN fraction is the number of galaxies with W1$-$W2 > 0.5 divided by the total number of target or control galaxies. For the WISE excess, the target samples and control pools are restricted to galaxies with WISE detections, even though the matching parameters are derived from SDSS spectra. The LERG fraction is the number of LERGs in each sample divided by the total number of target or control galaxies. For the LERG excess we require that the target and control pool galaxies are classified in the \citet{10.1111/j.1365-2966.2012.20414.x} catalog as either star forming galaxies, HERGs, or LERGs. All excesses are the ratio of the AGN fraction in the target sample and the AGN fraction in the relevant matched control sample.

\subsubsection{Optical AGN excess}
\label{Optical AGN excess}

The top panel of Figure~\ref{fig:exc-mz} shows the optical excesses for spectroscopic pairs (blue) in 8 bins of projected separation between 0$-$80$h_{70}^{-1}$ kpc, as well as for \citet{2013MNRAS.435.3627E} post-mergers (teal) and the visually confirmed post-merger sample (magenta). In matching controls for Figure~\ref{fig:exc-mz}, we reached the cap of 10 controls per target (see Section~\ref{AGN in post-mergers} above) in both post-merger samples, and in each bin of projected separation. We uncover a decreasing trend of optical AGN excess with increasing $r_{p}$, peaking with an excess of \textasciitilde1.7 for galaxy pairs separated by $\leq$ 10$h_{70}^{-1}$ kpc. Using a more generous optical AGN criterion (\citealp{2006MNRAS.371..972S}, which allows for the inclusion of composite galaxies with significant contributions from both star formation and AGN), and a different control matching algorithm (allowing for different numbers of controls to be selected for galaxies belonging to the same target sample), \citet{2011MNRAS.418.2043E} still reported quantitatively consistent pair-phase excesses, from \textasciitilde0$-$2.5 for galaxy pairs spanning separations of 80$-$0 $h_{70}^{-1}$ kpc. We recalculate the post-merger optical AGN excess in the \citet{2013MNRAS.435.3627E} sample in order to compare using our updated control-matching method, and find \textasciitilde2 times as many optical AGN compared to controls. We find that optical AGN are even more common in the visually confirmed post-merger sample, with an excess of 3.7.

\subsubsection{WISE AGN excess}
\label{WISE AGN excess}

There are 6,106 SDSS pairs, 364 visually confirmed post-mergers, and 78 \citet{2013MNRAS.435.3627E} post-mergers for which we can calculate a WISE mid-IR colour. The middle panel of Figure~\ref{fig:exc-mz} shows the excess of mid-IR AGN with W1$-$W2 > 0.5 in the same three main galaxy samples as in Section~\ref{Optical AGN excess} compared to matched control samples, which are themselves required to have WISE detections so that their WISE W1$-$W2 colours are calculable. Much like the optical AGN excess, we find a decreasing excess of mid-IR AGN in bins of $r_{p}$ with a peak value of \textasciitilde3.6 for the closest spectroscopic pairs. These results are statistically consistent with the pair phase W1$-$W2 > 0.5 results reported in \citet{2014MNRAS.441.1297S}, even though a different control matching methodology was used. Our visually confirmed post-merger sample shows a mid-IR AGN excess consistent with the closest pairs, but significantly below the \citet{2013MNRAS.435.3627E} sample, which contains 13.3 times as many WISE AGN per capita compared to the control sample. Our pair phase and visually confirmed post-merger sample excesses for both optical AGN and mid-IR AGN are in reasonable concordance (spanning \textasciitilde0$-$4 in the pair phase, and 3$-$4 in the post-mergers). This agreement supports the view that the optical and mid-IR AGN diagnostics are identifying different observational phenomena associated with the same SMBH engines.
\begin{figure}
\includegraphics[width=\columnwidth]{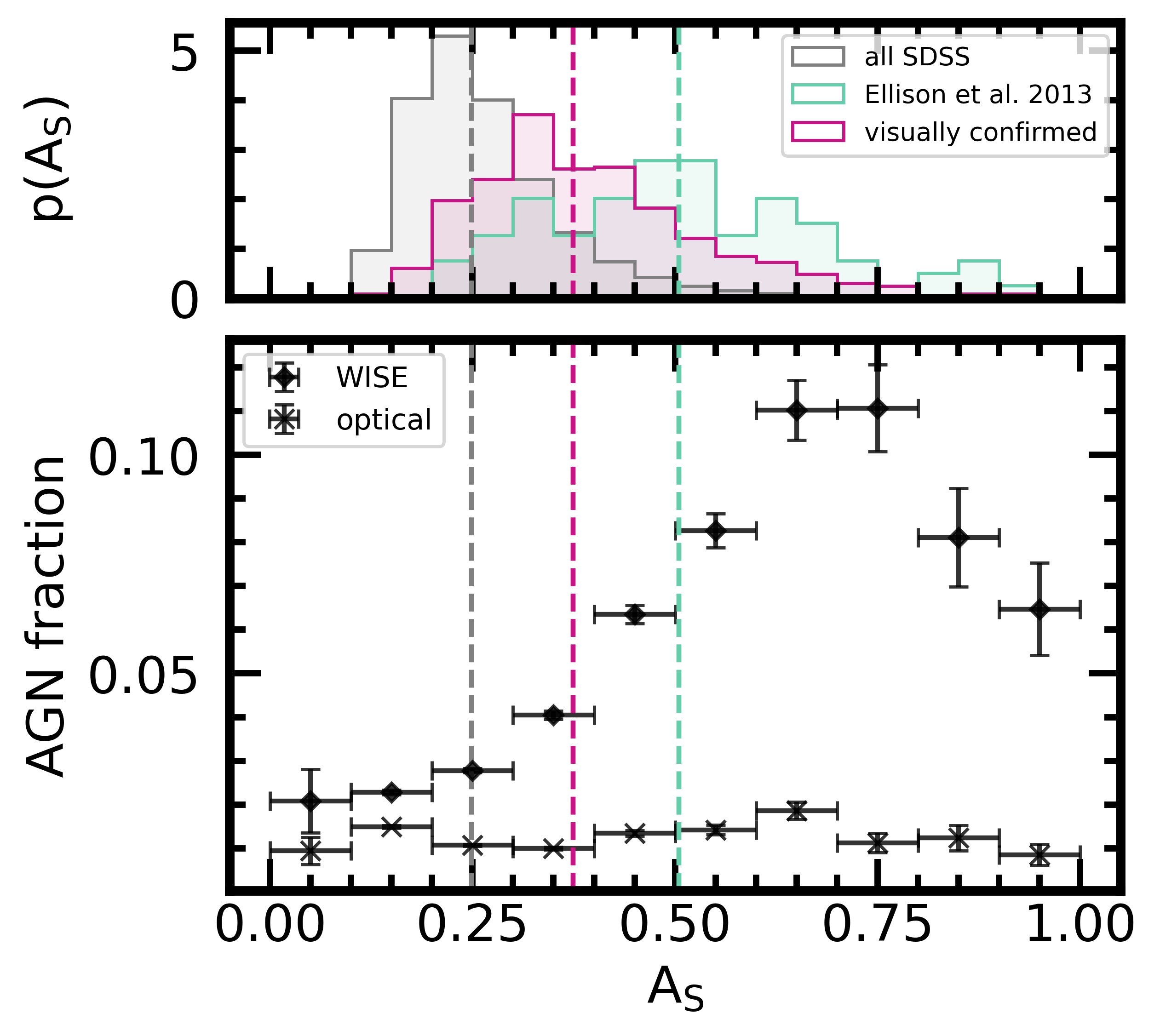}
\caption{Shape asymmetry statistics derived from SDSS imaging for SDSS galaxies with $z$ < 0.35 and $\mathrm{M_{\star}}$ > $\mathrm{10^{8.5}}$ \(\textup{M}_\odot\) (grey), the \citet{2013MNRAS.435.3627E} post-mergers (teal), and the visually confirmed post-mergers (magenta, top panel). The vertical bars spanning both panels represent the median of each distribution. While both more disturbed than the typical SDSS galaxy, the visually confirmed post-merger samples are less disturbed on average than the \citet{2013MNRAS.435.3627E} mergers, suggesting that a greater number of minor mergers are included. This would be a natural consequence of the CNN's inclusive simulated training set. The bottom panel shows the AGN fractions in the SDSS sample as identified by optical (X-markers) and mid-IR (diamonds) criteria, see Section~\ref{Methods} for details. The vertical errors are the binomial errors on each fraction $\sqrt{f(1-f)/N}$, and the horizontal errors are the bin widths. Since the hybrid method typically selects galaxies with smaller shape asymmetries, it also selects a proportionally lower fraction of dust-obscured AGN. This accounts for the discrepancy between the two post-merger data points in Figure~\ref{fig:exc-mz}.}
\label{fig:sa-diag}
\end{figure}

In order to investigate the discrepancies in optical and mid-IR AGN excesses between the visually confirmed post-mergers and the \citet{2013MNRAS.435.3627E} post-mergers, we investigate differences in the selection methods used for their identification. Because the CNN, Galaxy Zoo classifications, and expert visual classifications all rely on morphology, we posit that a morphological bias may be responsible. In order to investigate differences in sample morphology, we use shape asymmetry. Shape asymmetry is a non-parametric morphological measurement that takes the asymmetry of a binary mask that denotes a boundary between the galaxy and the background, detailed in \citet{2016MNRAS.456.3032P}. The binary mask used to measure shape asymmetry is generated following the 8-connected structure detection method described in \citet{2016MNRAS.456.3032P}, in which a galaxy image is smoothed using a 3$\times$3 running average filter, and pixels above a limiting surface brightness of approximately 24.7 mag arcsec$^{~-2}$, equivalent to one standard deviation above the typical sky noise level in SDSS imaging, are ascribed to the galaxy. It was developed for the purpose of automated merger identification because of its emphasis on the particular asymmetry of low-surface brightness features whose importance would be overlooked by a traditional asymmetry measurement. The merit of this metric for merger identification is explored in detail in \citet{2022arXiv220704152W}. By deploying shape asymmetry on galaxies whose post-merger status is already confirmed, shape asymmetry instead measures the morphological strength of the post-merger. Tidally disturbed post-mergers who have experienced dramatic, major mergers are more likely to have more extended morphologies, and higher shape asymmetries compared to those whose mergers have been relatively tame. Shape asymmetry has also been shown to fade in the several hundred Myr that follow coalescence \citep{2016MNRAS.456.3032P}, but so too has it been shown that the W1$-$W2 colour decreases after coalescence \citep{2018MNRAS.478.3056B}. Shape asymmetry therefore also includes information about the recency of the merger, along with the initial intensity of the morphological disruption.  

Figure~\ref{fig:sa-diag} investigates the shape asymmetry demographics of our two post-merger samples. The top panel shows the normalized distributions of shape asymmetry derived from SDSS imaging for the entire SDSS DR7 galaxy population with spectroscopic redshifts < 0.35, and masses > $\mathrm{10^{8.5}}$ \(\textup{M}_\odot\) (grey, representing the area of $\mathrm{M_{\star}-}z$ parameter space encompassing both the \citet{2013MNRAS.435.3627E} and visually-confirmed post-merger samples, see Figure~\ref{fig:m-z-dist}), the \citet{2013MNRAS.435.3627E} post-merger sample (teal), and the new visually confirmed sample (magenta). It is important to note that shape asymmetry does not trend strongly with either stellar mass or redshift, and that the qualitative results of our shape asymmetry study do not change when we compare our merger samples to their matched control galaxies from Sections~\ref{Optical AGN excess} and~\ref{WISE AGN excess} instead of the SDSS parent sample used here. The median SDSS-derived shape asymmetry of each sample is plotted over both panels as a dashed line of the same colour. Shape asymmetries derived from CFIS $r$-band imaging are available for the visually confirmed post-mergers, but we present only SDSS shape asymmetries in order to allow for direct comparison to the \citet{2013MNRAS.435.3627E} sample, which does not appear in full in CFIS. Note that the shapes of the visually confirmed post-merger sample, while of course more asymmetrical than SDSS in general, are significantly less disturbed than the \citet{2013MNRAS.435.3627E} post-mergers, with $\bar \Delta A_{S}$ \textasciitilde 0.13. We posit that the typical difference in SDSS-derived shape asymmetry between the two post-merger samples is the result of the fact that the \citet{2013MNRAS.435.3627E} post-mergers were identified by strictly visual means in shallow imaging. Conversely, the visually confirmed post-mergers were first identified by a CNN trained on CFIS-depth simulated imaging of post-mergers with mass ratios as small as 10:1. It is therefore plausible that a number of relatively minor post-mergers were preserved by the CNN and confirmed during visual classification.

The bottom panel of Figure~\ref{fig:sa-diag} shows the local optical and WISE AGN fractions of the SDSS parent sample (with $z$ < 0.35 and $\mathrm{M_{\star}}$ > $\mathrm{10^{8.5}}$ \(\textup{M}_\odot\)) in 10 bins of $A_{S}$ between 0 and 1. We find that the optical AGN fraction is generally low and consistent with increasing shape asymmetry. While the WISE AGN fraction does not trend monotonically with $A_{S}$, the data show that more morphologically disturbed galaxies in SDSS are indeed more likely to host an AGN that is identifiable by its mid-IR colour. These results indicate that the degree of disturbance is unlikely to have a strong impact on the optical AGN fraction, but this is not true for WISE AGN, since more disturbed galaxies typically have higher mid-IR AGN fractions. Consequently, the \citet{2013MNRAS.435.3627E} sample is more likely to contain highly disturbed post-mergers, which host proportionally more dust-obscured AGN and a consistent number of optical AGN, while the visually confirmed sample is more inclusive of less-disturbed mergers, which are less likely to host mid-IR AGN.

If the degree of morphological disturbance is responsible for the difference in mid-IR AGN excess between the two post-merger samples, a subset of the visually confirmed post-mergers with the same SDSS shape asymmetry demographics as the \citet{2013MNRAS.435.3627E} post-mergers ought to exhibit an excess that is in better agreement. In order to test this hypothesis, we match exactly one galaxy (without replacement) from the visually confirmed post-merger sample to each \citet{2013MNRAS.435.3627E} post-merger on shape asymmetry. Where multiple visually confirmed post-mergers have shape asymmetries within $\pm$5\% of an \citet{2013MNRAS.435.3627E} post-merger, we select the single best match. Where there are no matches, we grow the tolerance from 5\% until a single match can be found. Of the 85 galaxies in the \citet{2013MNRAS.435.3627E} sample with shape asymmetries available, 82 have a match within 5\% of their shape asymmetry in the visually confirmed sample. The remaining 3 galaxies require 2, 3, and 5 growths, respectively; they are included for completeness but their exclusion does not affect our results. The shape asymmetry-matched visually confirmed post-mergers have an optical AGN excess consistent with the visually confirmed sample taken as a whole, but their mid-IR AGN excess is increased from 3.6 up to 6.5. While still not in perfect agreement with the \citet{2013MNRAS.435.3627E} sample, this experiment confirms that the degree of morphological disturbance ($\bar \Delta A_{S}$ \textasciitilde 0.13) and sample selection are linked to an increased likelihood of a mid-IR AGN detection. This result is consistent with the physical narrative presented in \citet{2022arXiv220500567Y}, wherein rapidly accreting AGN in extremely recent (within \textasciitilde4Myr) post-coalescence systems are more likely to be observed as dust-obscured galaxies (DOGs) on account of central and/or galaxy-scale dispersion of dust from the progenitor galaxies. We posit that the longevity and intensity of the dust obscuration may scale with the dynamic intensity of the merger. As a result, the specific method used to identify mergers based on their morphology has a significant impact on the quantitative excesses we calculate.

\subsubsection{LERGs in post-mergers}
\label{LERGs in post-mergers}

The bottom panel of Figure~\ref{fig:exc-mz} investigates the role of mergers in triggering LERGs. In our mass- and redshift-matched study, we find that LERGs are no more likely to exist in our post-merger or pair samples than in controls. This result is qualitatively discrepant with the \citet{2014ApJ...785...66P} finding that galaxies hosting radio AGN have a 50\% excess in the number of satellites, and the link between tidal forces associated with pair phase interactions and LERG incidence suggested by \citet{10.1093/mnras/sts675}. \citet{2015MNRAS.451L..35E} find a modest pair phase LERG excess of 3.8$\pm$0.4, and a small excess of \textasciitilde4$\pm$2 (nearly consistent with unity as well) in the post-merger phase when they match controls on stellar mass and redshift. The lack of an elevated LERG incidence rate in post-mergers or galaxy pairs in this work is therefore in mild tension with the literature, even though the conditions in a post-merger system are certainly not required for the triggering of LERGs.

\subsection{Optical AGN accretion rate enhancements}
\label{Optical AGN accretion rate enhancements}

In addition to the initial triggering of AGN, we can use our post-merger sample to determine typical merger-induced accretion rate enhancements in optically-identified AGN, using [OIII] luminosity as a proxy for accretion rate (see also \citealp{2003MNRAS.346.1055K}; \citealp{2004MNRAS.351.1151B}; \citealp{2009ApJ...695L.130C}; \citealp{2012ApJ..745...94L}). As stated in Section~\ref{Optical AGN}, we again use a S/N criterion of at least 5 for the four BPT emission lines, and explicitly disallow LINERS using the [SII] and [OI] criteria of \citet{2006MNRAS.372..961K}. Because we are computing a luminosity enhancement, we require an ensemble of at least 5 AGN controls for each target galaxy in order to compare the luminosity of each AGN post-merger or pair to a group of non-merger or non-pair counterparts. Rather than finding the nearest controls in parameter space, we set initial tolerances of $\pm0.1$ dex in M$\star$ and $\pm0.05$ in $z$. In practice, all of our target galaxies (pairs and post-mergers alike) find at least 5 controls without any growths in parameter space. 31 post-mergers from the visually confirmed sample, 3 from the \citet{2013MNRAS.435.3627E} sample, and 263 SDSS pairs in bins of separation between 0$-$80$h_{70}^{-1}$ kpc with mass ratios $0.1 \leq$ M$_{1}$/M$_{2} \leq 10$ selected from the \citet{2016MNRAS.461.2589P} catalog with non-LINER optical AGN and S/N of at least 5 on all emission lines used for placement on the BPT diagram are ultimately included. The [OIII] luminosity enhancement, $\Delta$log(L$_\mathrm{[OIII]}$), is calculated as the difference between the logged [OIII] luminosity (in units of $\mathrm{erg~s^{-1}}$) of the target galaxy and the median logged luminosity of the control ensemble, and hence captures the typical accretion rate difference between AGN triggered my mergers and secular AGN.

\begin{figure}
\includegraphics[width=\columnwidth]{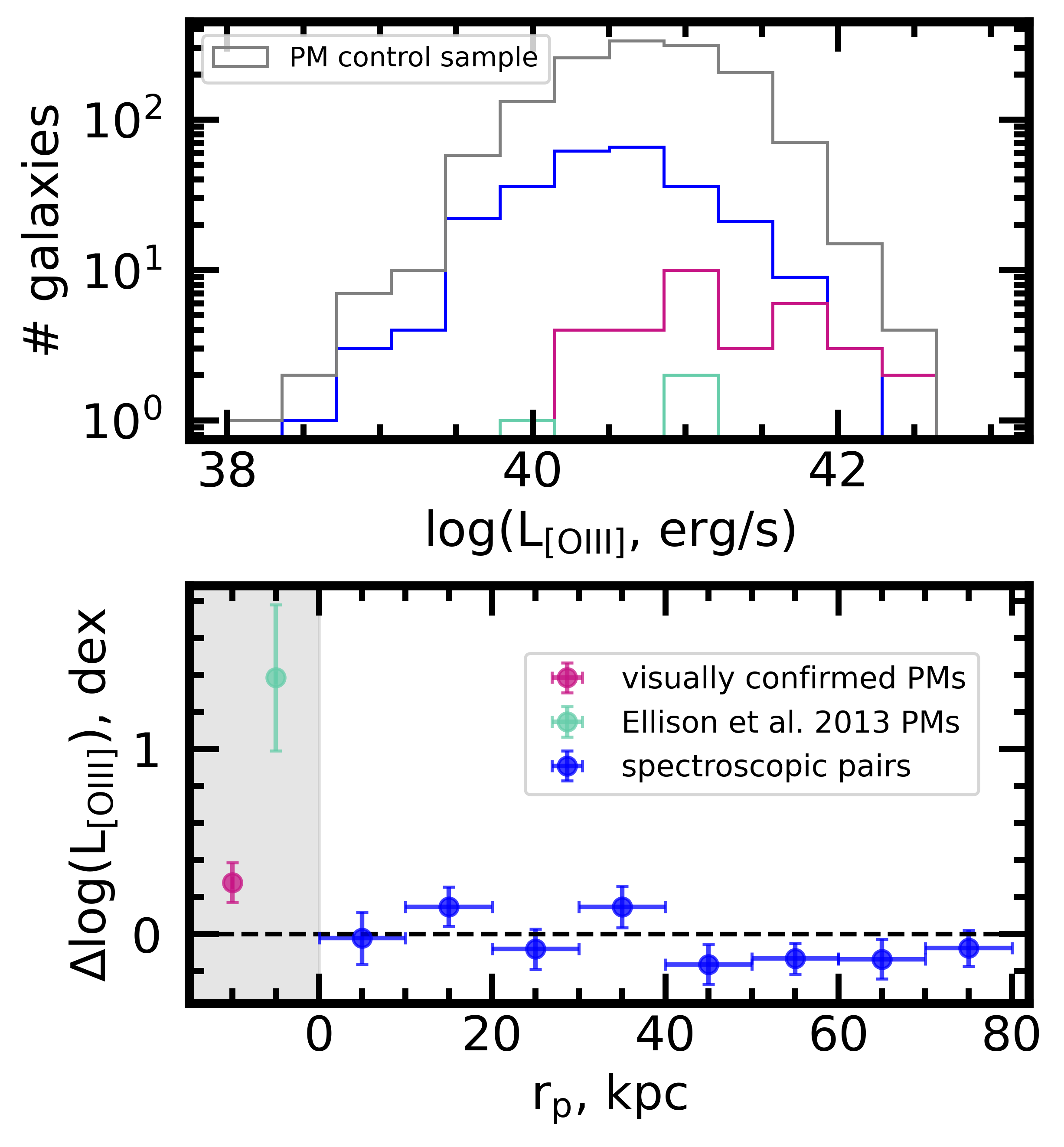}
\caption{[OIII] luminosities and luminosity enhancements in post-mergers and pairs. The top panel shows log-scale [OIII] luminosity histograms for optical AGN in the the galaxy pair sample described in Section~\ref{Galaxy pairs} (blue), the visually confirmed post-merger sample (magenta), the \citet{2013MNRAS.435.3627E} post-merger sample (teal), and the optical AGN control pool for the visually confirmed post-merger sample (grey). The bottom panel shows $\Delta$log(L$_\mathrm{[OIII]}$) for the same three target samples (galaxy pairs, visually confirmed post-mergers, and \citealp{2013MNRAS.435.3627E} post-mergers). Vertical error bars are the statistical error on the median, $1.253\sigma/\sqrt{N}$, and horizontal error bars are the bin widths. We find enhancements approximately consistent with zero in the pair phase, with some small local suppressions past 40$h_{70}^{-1}$ kpc. In both post-merger samples, we find significant positive excesses. Post-mergers in the visually confirmed sample are \textasciitilde2 times as luminous in [OIII] as their non-post-merger controls.}
\label{fig:O3}
\end{figure}

Figure~\ref{fig:O3}, which shows L$_\mathrm{[OIII]}$ (top panel) and $\Delta$log(L$_\mathrm{[OIII]}$) (bottom panel) for optical AGN in spectroscopic pairs, the \citet{2013MNRAS.435.3627E} post-mergers, and the visually confirmed post-merger sample, suggests that accretion rates in AGN hosted by galaxies with a close companion are consistent with isolated AGN. Conversely, both post-merger samples are significantly enhanced compared to the matched non-merger AGN control ensembles. The visually confirmed post-merger sample is in fact \textasciitilde2 times as luminous in [OIII] on average compared to non-post-merger controls. The accretion rate enhancements for optical AGN ushered in by coalescence therefore appear to be significant, while those produced by pair-phase interactions may just as likely be produced by secular or ambient processes. A higher positive enhancement of \textasciitilde1.4 dex is recovered for the three post-mergers that remain after applying our quality control cuts to the \citet{2013MNRAS.435.3627E} post-mergers. While the sample size does not invite extensive interpretation, it is possible that the increased morphological disturbance that is typical of the \citet{2013MNRAS.435.3627E} sample is linked to more rapid gas infall and elevated accretion rates in these systems.

The relationship between star formation rate and L$_\mathrm{[OIII]}$ is of order unity (with significant scatter) in BPT star forming galaxies beneath the \citet{2003MNRAS.346.1055K} criterion with S/N$>$5. Therefore, in order to achieve similar $\Delta$log(L$_\mathrm{[OIII]}$), star forming galaxies would need to have approximately doubled star formation rates (SFRs). Still, since we select galaxies whose nebular emission could not plausibly be produced by star formation alone, the [OIII] luminosity enhancements uncovered in this Section are primarily indicative of SMBH accretion-driven ionization.

The result does not appear to be an effect of the SDSS fiber aperture on the \citet{2013MNRAS.435.3627E} sample, which lies at low-$z$ relative to the visually confirmed sample (see Figure~\ref{fig:m-z-dist}) as we find no correlation of $\Delta$log(L$_\mathrm{[OIII]}$) with $z$ for individual galaxies in the visually confirmed sample. In the sample of \citet{2001ApJ...556..121K} AGN with S/N $\geq5$ for the four BPT emission lines and stellar masses between $\mathrm{10^{10-12}}$ \(\textup{M}_\odot\), the median measured [OIII] luminosity actually increases from $10^{39.2}$ to $10^{41.6}$ between 0 $\leq z \leq$ 0.25. The fiber aperture effect therefore gives rise to higher [OIII] luminosities at higher $z$, and moreover, our control-matching methodology accounts for systematic changes in [OIII] luminosity with stellar mass and redshift. Aperture effects (or more broadly, any redshift or stellar mass effects) are not responsible for the difference in $\Delta$log(L$_\mathrm{[OIII]}$) calculated between the \citet{2013MNRAS.435.3627E} and visually confirmed post-mergers.

\section{Summary}
\label{Summary}

In this work, we have used the CNN-identified and visually confirmed post-merger sample introduced in \citet{2022MNRAS.514.3294B} to study the triggering and accretion of supermassive black holes in post-merger galaxies. We also offer pair phase results in order to contextualize the post-merger results within the merger sequence. We match control (non-post-merger or non-pair) galaxies on M$\star$ and $z$ to our target (either post-merger or pair) galaxies in order to study the AGN excess --- that is, the ratio of the AGN fractions in the target sample and the control sample --- using optical narrow-line region (via SDSS, the BPT diagram, and the \citealp{2001ApJ...556..121K} AGN selection), mid-IR (via WISE and the colour criterion W1$-$W2 > 0.5 used by \citealp{2014MNRAS.441.1297S} to select dust-obscured AGN), and LERG classifications (capturing the low-excitation, isotropic SMBH accretion mode, as determined by \citealp{10.1111/j.1365-2966.2012.20414.x}). We report the following:
\begin{itemize}
\item We find that optical and mid-IR AGN excesses in the pair sample increase from \textasciitilde1 (i.e., no excess) to \textasciitilde2$-$4 as their projected separations decrease from $r_{p}$\textasciitilde80 $h_{70}^{-1}$ kpc down to zero. We find that galaxy pairs and mergers do not preferentially exhibit the characteristics of LERGs, in mild tension with the literature (Figure~\ref{fig:exc-mz}).
\medskip
\item We find optical and mid-IR AGN excesses in a new sample of visually confirmed post-mergers of \textasciitilde$3-4$ over controls matched on mass and redshift, suggesting that the nuclear conditions ushered in by post-mergers increase the likelihood that an energetic AGN will be triggered.
\medskip
\item We find a significant connection between high shape asymmetry derived from SDSS imaging (roughly analogous to the degree of merger disturbance) and a galaxy's likelihood to host a dust-obscured AGN in an inclusive sample of SDSS galaxies. This is most likely related to the tendency of mergers to disperse the central gas and dust belonging to their participant galaxies. This connection contributes to the quantitative differences in the optical and mid-IR AGN excesses of our visually confirmed merger sample and that of \citet{2013MNRAS.435.3627E}, which is composed of more visually dramatic merger examples with higher shape asymmetries on average (Figure~\ref{fig:sa-diag}).
%\medskip
%\item Including additional control-matching parameters ($\delta_{5}$ (environment density), $D_{4000}$, (the strength of the 4000\AA~break), and $\mathrm{M_{Halo}}$) in our excess experiment does affect the results. Our pair phase optical and mid-IR trends remain significant, while our optical and mid-IR post-merger excesses decrease (down to 1, or no excess, in the case of the optical AGN excess). We suggest that this attenuation occurs because it is the central gas reservoir brought in by a merger, rather than the merger itself, that triggers the AGN, and $D_{4000}$ and $\mathrm{M_{Halo}}$ indirectly control for the presence of a gas reservoir. The LERG excess, meanwhile, is effectively erased in both the pair and post-merger phase when these factors are considered, suggesting that LERGs are not preferentially produced by galaxy mergers, but rather by ambient and secular conditions that happen to correlate with merger status. While the other two additional parameters do affect small changes on the quantitative LERG excesses we calculate, $D_{4000}$ is largely responsible for the collapse of the LERG excess (Figure~\ref{fig:exc-deet}). This result is consistent with that in \citet{2015MNRAS.451L..35E}.
\medskip
\item Following a number of efforts in the literature (see Section~\ref{Optical AGN accretion rate enhancements}), we use [OIII] luminosity as a proxy for SMBH accretion rate in a sample of optical AGN. We find that optical AGN hosted by interacting galaxy pairs are not preferentially enhanced in their accretion rates (as measured by L$_\mathrm{[OIII]}$) compared with secularly driven AGN in isolated galaxies. Conversely, we find our visually confirmed post-merger sample to be \textasciitilde2 times as bright in [OIII] than the AGN in isolated galaxies. This suggests that the typical accretion rate enhancements produced during the pair phase of the merger sequence are just as likely to be produced by secular or ambient processes (e.g. halo gas accretion, secular gas accretion from stellar winds or supernovae), while the post-merger phase produces significant accretion rate enhancements. This result, as well as our excess results, support an important role for the post-merger epoch in triggering and growing the SMBHs residing at the core of every galaxy (Figure~\ref{fig:O3}).
\end{itemize}

In addition to the above detailed census of AGN in post-coalescence galaxies, we can also revisit the merit of a hybrid (CNN plus human visual classification) post-merger identification framework, which has allowed us to improve on the statistics of literature studies of post-mergers, and propose revisions to other results whose quantities were more heavily influenced by the selection functions of their merger identification method. Because the visually confirmed CFIS merger sample is biased only by the training of the CNN and the decisions of the visual classification team, we believe the galaxies themselves (catalogued in \citealp{2022MNRAS.514.3294B}) will continue to provide value in the form of subsequent cross-survey characterization. Moreover, the hybrid classification framework itself shows promise for future questions in astronomy surrounding rare and elusive observational phenomena.

\section*{Acknowledgements}
\label{Acknowledgements}

The work detailed above was conducted at the University of Victoria in Victoria, British Columbia, as well as in the Township of Esquimalt in Greater Victoria. We acknowledge with respect the Lekwungen peoples on whose unceded traditional territory the university stands, and the Songhees, Esquimalt and $\mathrm{\ubar{\mathrm{W}}S\acute{A}NE\acute{C}}$ peoples who have stewarded the land for centuries and continue to do so today.

CFIS is conducted at the Canada-France-Hawaii Telescope on Maunakea in Hawaii. We also recognize and acknowledge with respect the cultural importance of the summit of Maunakea to a broad cross section of the Native Hawaiian community.

We thank Samir Salim, Christopher Agostino, and Connor Bottrell for their indespensible feedback on this work.

This work is based on data obtained as part of the Canada-France Imaging Survey, a CFHT large program of the National Research Council of Canada and the French Centre National de la Recherche Scientifique, and on observations obtained with MegaPrime/MegaCam, a joint project of CFHT and CEA Saclay, at the Canada-France-Hawaii Telescope (CFHT) which is operated by the National Research Council (NRC) of Canada, the Institut National des Science de l’Univers (INSU) of the Centre National de la Recherche Scientifique (CNRS) of France, and the University of Hawaii. This research used the facilities of the Canadian Astronomy Data Centre operated by the National Research Council of Canada with the support of the Canadian Space Agency.

Data from the IllustrisTNG simulations are integral to this work. We thank the Illustris Collaboration for making these data available to the public.

Funding for the SDSS and SDSS-II has been provided by the Alfred P. Sloan Foundation, the Participating Institutions, the National Science Foundation, the U.S. Department of Energy, the National Aeronautics and Space Administration, the Japanese Monbukagakusho, the Max Planck Society, and the Higher Education Funding Council for England. The SDSS Web Site is http://www.sdss.org/. The SDSS is managed by the Astrophysical Research Consortium for the Participating Institutions. The Participating Institutions are the American Museum of Natural History, Astrophysical Institute Potsdam, University of Basel, University of Cambridge, Case Western Reserve University, University of Chicago, Drexel University, Fermilab, the Institute for Advanced Study, the Japan Participation Group, Johns Hopkins University, the Joint Institute for Nuclear Astrophysics, the Kavli Institute for Particle Astrophysics and Cosmology, the Korean Scientist Group, the Chinese Academy of Sciences (LAMOST), Los Alamos National Laboratory, the Max-Planck-Institute for Astronomy (MPIA), the Max-Planck-Institute for Astrophysics (MPA), New Mexico State University, Ohio State University, University of Pittsburgh, University of Portsmouth, Princeton University, the United States Naval Observatory, and the University of Washington.

This research was enabled, in part, by the computing resources provided by Compute Canada.

%%%%%%%%%%%%%%%%%%%%%%%%%%%%%%%%%%%%%%%%%%%%%%%%%%
\section*{Data Availability}
\label{Data Availability}

Simulation data from TNG100-1 used in the generation of training images for this work are openly available on the IllustrisTNG website, at tng-project.org/data. Template versions of \textsc{RealSim} and \textsc{RealSim-CFIS}, developed by Connor Bottrell with modifications by RWB are publicly available via GitHub at github.com/cbottrell/RealSim and github.com/cbottrell/RealSim-CFIS. Specific image training data used to develop the findings of this study are available by request from RWB.

This publication makes use of data products from the Wide-field Infrared Survey Explorer, which is a joint project of the University of California, Los Angeles, and the Jet Propulsion Laboratory / California Institute of Technology, funded by the National Aeronautics and Space Administration. Data from the Sloan Digital Sky Survey are available at sdss.org.

The public visually confirmed post-merger catalog is available via MNRAS as a digital resource along with \citet{2022MNRAS.514.3294B}.

The Canada France Imaging Survey is a legacy survey for the Canadian and French communities. All data are proprietary within the Canadian and French communities until Aug. 1st, 2023 when the individual frames will become available to the worldwide community. Advanced curated CFIS data products will be made available to the world as part of UNIONS public data releases. 



\bibliographystyle{mnras}
\bibliography{bib_01} % if your bibtex file is called example.bib

%\begin{thebibliography}{99}
%\end{thebibliography}

%%%%%%%%%%%%%%%%%%%%%%%%%%%%%%%%%%%%%%%%%%%%%%%%%%

%%%%%%%%%%%%%%%%% APPENDICES %%%%%%%%%%%%%%%%%%%%%

%\appendix

%%%%%%%%%%%%%%%%%%%%%%%%%%%%%%%%%%%%%%%%%%%%%%%%%%

% Don't change these lines
\bsp	% typesetting comment
\label{lastpage}
\end{document}